\documentstyle[aps,prd,eqsecnum,twocolumn,epsf]{revtex}

\newcommand{\singlefig}[2]{
\begin{center}
\begin{minipage}{#1}
\epsfxsize=#1
\epsffile{#2}
\end{minipage}
\end{center}}
\newenvironment{figcaption}[2]{
 \vspace{0.3cm}
 \refstepcounter{figure}
 \label{#1}
 \begin{center}
 \begin{minipage}{#2}
 \begingroup \small FIG. \thefigure: }{
 \endgroup
 \end{minipage}
 \end{center}}
\newcommand{\gsim}{\mbox{\raisebox{-1.ex}{$\stackrel
     {\textstyle>}{\textstyle\sim}$}}}
\newcommand{\lsim}{\mbox{\raisebox{-1.ex}{$\stackrel
     {\textstyle<}{\textstyle \sim}$}}}
\newcommand{\square}{\kern1pt\vbox{\hrule height
1.2pt\hbox{\vrule width 1.2pt\hskip 3pt
   \vbox{\vskip 6pt}\hskip 3pt\vrule width 0.6pt}\hrule
height 0.6pt}\kern1pt}

\begin{document}
\draft \twocolumn[\hsize\textwidth\columnwidth\hsize\csname
@twocolumnfalse\endcsname

\title{A new twist to preheating}
\author{Shinji Tsujikawa$^*$ and Bruce A. Bassett$^{\P}$}
\address{$^*$ Department of Physics, Waseda University,
Shinjuku-ku, Tokyo 169-8555, Japan\\[.3em]}
\address{$^{\P}$ Relativity and Cosmology Group (RCG), University of
Portsmouth, Mercantile House, Portsmouth,  PO1 2EG, England}
\date{\today}
\maketitle
\begin{abstract}
Metric perturbations typically strengthen field resonances during
preheating. In contrast we present a model in which the
super-Hubble field resonances are completely {\em suppressed} when
metric perturbations are included. The model is the nonminimal
Fakir-Unruh scenario which is exactly solvable in the long-wavelength
limit when metric perturbations are included, but exhibits
exponential growth of super-Hubble modes in their absence. 
This gravitationally enhanced
integrability is exceptional, both for its rarity and for the power with
which it illustrates the importance of including metric
perturbations in consistent studies of preheating. We conjecture a
no-go result -  there exists no {\em single-field} model with
growth of cosmologically-relevant metric perturbations during 
preheating.
\end{abstract}

\vskip 1pc 
 ]

\section{Introduction}                           %

Gravity has persistently proven the most difficult of the forces
of Nature to bring under theoretical control, both at the
classical and quantum levels. This is due, in part, to the extreme
non-linear and mixed hyperbolic-elliptic  nature of the  field
equations. Further, the dimensionful  gravitational coupling
constant, $G = m_{\rm pl}^{-2}$,  leads to strong
non-renormalizability of even the linearized, quantum, theory in
four dimensions.

The semi-classical approximation was developed in partial response
to these obstacles. While there are strong indications that large
regions of the semi-classical solution space may be spurious
\cite{ash}, a consistent cosmology of inflation in the
post-Planckian universe has emerged. This  appears to lie
 in the semi-classical domain with  gravity playing a
gentle, domesticated,  r\^ole of quiet order within a highly
symmetric universe.

Preheating, in contrast, is  a violent, non-equilibrium and
anarchic epoch \cite{pre1,pre2} which may have linked this Utopian
regime to the older, radiation-dominated,  universe. The rapid
transfer of energy between fields during preheating can excite the
tiny metric perturbations produced during inflation, even on
cosmological scales and may lead to unbridled growth and
non-linearity \cite{mpre2}. This threatens to take the system away
from the safety of the semi-classical approximation \cite{mpre4}.

In these multi-field cases, neglecting metric perturbations can be
a grave mistake. Nevertheless, a belief has developed that metric
perturbations are unimportant for understanding the super-Hubble
evolution of a single scalar field \cite{mpre1}. This is true in
the minimally coupled case \cite{MFB}, but certainly {\em not}
true in the {\em non-minimally} coupled case we discuss below.


Since the non-minimal case was the only single field model in which
super-Hubble resonances might plausibly have existed, we are lead
to conjecture that there exist no single-field models with
super-Hubble metric resonances.


It was pointed out by Fakir and Unruh\cite{FU} that the
fine-tuning of $\lambda$, which plagues the minimal
$V(\phi)=\lambda\phi^4/4$ chaotic inflation model, is absent 
when $\xi$ is very large and negative (see also \cite{SBB,FM}).
Recently Tsujikawa {\it et al.}\cite{nonminimalpre} showed that
long-wave modes of the inflaton in this model are resonantly
enhanced during preheating for $\xi \ll -1$,  {\em when  metric
perturbations are neglected}, due to the oscillations of the Ricci
scalar \cite{nonminimalpre2}. Naively one might expect this
geometric preheating to cause super-Hubble modes of the
gauge-invariant metric perturbations to grow too \cite{mpre4}. In
fact, we will show that including metric perturbations has a more
surprising effect, namely to completely {\em remove} the
super-Hubble resonances.

 \section{The model and analytical estimates}
Consider the inflaton, $\phi$, coupled non-minimally to the
spacetime curvature $R$ with Lagrangian density:
\begin{eqnarray}
 {\cal L}= \sqrt{-g} \left[ \frac{1}{2\kappa^2}R
   -\frac{1}{2}(\nabla \phi)^2
   -V(\phi)
   -\frac12 \xi R \phi^2
    \right],
\label{B1}
\end{eqnarray}
with $G \equiv \kappa^{2}/8\pi =m_{\rm pl}^{-2} $ the
gravitational coupling constant, and $\xi$  the non-minimal
coupling. For a massive inflaton, $|\xi|~\lsim~10^{-3}$ is
required  for sufficient inflation. On the other hand, in the
massless case with self-interaction
\begin{eqnarray}
V(\phi)=\lambda \phi^4/4,
\label{B2}
\end{eqnarray}
such a constraint is absent for negative $\xi$.
In this paper, we consider the Fakir-Unruh scenario\cite{FU}
where the potential is described by $(\ref{B2})$ with
negative $\xi$.\footnote{We consider nagative
$\xi$ since large positive $\xi$ leads to a repulsive effective
gravitational coupling.} 

We choose a flat FLRW background and consider the perturbed metric
in the longitudinal gauge:
\begin{eqnarray}
ds^2=-(1+2\Phi)dt^2
+a^2(1-2\Psi)\delta_{ij} dx^i dx^j,
\label{B3}
\end{eqnarray}
where $\Phi$ and $\Psi$ are gauge-invariant potentials\cite{MFB}.
 We decompose the inflaton field as $\phi(t,{\bf
x})=\phi_0(t)+\delta\phi(t,{\bf x})$, where $\phi_0$ is the
homogeneous condensate and $\delta\phi$ is the gauge-invariant
fluctuation. The Fourier modes of the first order perturbed
Einstein equations are then as\cite{Hwang}
\begin{eqnarray}
\Phi_k=\Psi_k-\delta F_k/F
\label{B4}
\end{eqnarray}
\begin{eqnarray}
\dot{\Psi}_k + \left(H+\frac{\dot{F}}{F}\right) \Phi_k 
=\frac{\kappa^2}{2F}\left(\dot{\phi}_0\delta\phi_k+\delta \dot{F}_k
-H\delta F_k \right),
\label{B5}
\end{eqnarray}
\begin{eqnarray}
& &\ddot{\Psi}_k + H\dot{\Psi}_k+\left(H+\frac{\dot{F}}{2F}\right)
\left(2\dot{\Psi}_k+\dot{\Phi}_k\right)+\frac{\kappa^2 \lambda
\phi_0^4}{4F}\Phi_k \nonumber \\ 
&=& \frac{1}{2F} \Biggl[\delta \ddot{F}_k+2H\delta\dot{F}_k-
\left(\frac{\kappa^2 \lambda \phi_0^4}{4}-
\frac12 \ddot{F}-\frac52 H \dot{F}_k\right)
\frac{\delta F_k}{F} \nonumber \\ 
&+& \kappa^2 \left\{ \dot{\phi}_0 \delta \dot{\phi}_k
-(\xi R+\lambda\phi_0^2)\phi_0\delta\phi_k \right\}
\Biggr],
\label{B60}
\end{eqnarray}
\begin{eqnarray}
 \delta \ddot{\phi}_k &+& 3H\delta \dot{\phi}_k+\left(
k^2/a^2+ 3\lambda\phi_0^2+\xi R \right) \delta \phi_k
\nonumber \\ &=& \dot{\phi}_0
(\dot{\Phi}_k+6H\Phi_k+3\dot{\Psi}_k)
 +2\ddot{\phi}_0 \Phi_k
+ 2\xi\phi_0 \nonumber \\
&\times& \biggl[3(2\dot{H}\Phi_k+H\dot{\Phi}_k+\ddot{\Psi}_k
+4H^2\Phi_k+4H\dot{\Psi}_k) \nonumber \\
&-& k^2/a^2(\Phi_k-2\Psi_k) \biggr],
\label{B6}
\end{eqnarray}
where $F=1-\xi\kappa^2\phi_0^2$, $\delta 
F_k=-2\xi\kappa^2\phi_0\delta\phi_k$, 
and $H \equiv \dot{a}/{a}$ is the Hubble expansion rate. 
{}From Eq.~$(\ref{B4})$ it is clear that $\Phi_k$ and $\Psi_k$ do not
coincide in the non-minimally coupled case, due to the
non-vanishing anisotropic stress in this case. We include the
backreaction of inflaton fluctuations on the  background equations
for the scale factor and the condensate $\phi_0$, via the Hartree
approximation \cite{nonminimalpre,KT1,structure}, yielding
\begin{eqnarray}
 H^2 &=&
   \frac{\kappa^2}{3(1-\xi\kappa^2\langle\phi^2\rangle)}
     \Biggl[ \frac12 (\dot{\phi}_0^2+
    \langle \delta \dot{\phi}^2 \rangle)+
    \left( \frac12 -2\xi \right) \langle \delta\phi'^2 \rangle
    \nonumber \\
  &&  +\frac14 \lambda(\phi_0^4+6\phi_0^2 \langle \delta\phi^2 \rangle
    +3\langle \delta\phi^2 \rangle^2)
    \nonumber \\
  && +2\xi \{3H(\phi_0 \dot{\phi}_0+\langle \delta\phi \delta
\dot{\phi} \rangle) -\langle \delta\phi \delta \phi'' \rangle \}
\Biggr],
\label{B7}
\end{eqnarray}
\begin{eqnarray}
\ddot{\phi}_0 +3H \dot{\phi}_0 +\lambda \phi_0 (\phi_0^2+3 \langle
\delta\phi^2 \rangle) +\xi R \phi_0 =0\,, \label{B8}
\end{eqnarray}
where $R$ is the scalar curvature, given by
\begin{eqnarray}
 R &=&
   \frac{\kappa^2}{1-\xi\kappa^2\langle\phi^2\rangle}
     \Biggl[ (1-6\xi)(-\dot{\phi}_0^2 -\langle \delta\dot{\phi}^2 \rangle
    +\langle \delta\phi'^2 \rangle)
    \nonumber \\
&&    + \lambda(\phi_0^4+6\phi_0^2 \langle \delta\phi^2 \rangle
    +3\langle \delta\phi^2 \rangle^2)
    +6\xi \{ \phi_0 \ddot{\phi}_0
 \nonumber \\
&& +\langle \delta\phi \delta \ddot{\phi} \rangle+3H(\phi_0
\dot{\phi}_0+\langle \delta\phi \delta \dot{\phi} \rangle)
-\langle \delta\phi \delta \phi'' \rangle \} \Biggr].
\label{B9}
\end{eqnarray}
The expectation values of $\delta \phi^2$ and $\phi^2$ are defined as
\begin{eqnarray}
\langle \delta \phi^2 \rangle &=& \frac1{2\pi^2} \int k^2
|\delta \phi_k|^2 dk,\\
\langle \phi^2 \rangle &=& \phi_0^2
+\langle \delta \phi^2 \rangle.
\label{B10}
\end{eqnarray}
Since the growth of the variance $\langle \delta \phi^2 \rangle$
affects the evolution of the fluctuations, we change  $\phi_0^2$
in Eqs.~$(\ref{B4})$-$(\ref{B6})$ to $\langle \phi^2 \rangle$. In
principle we should also consider the backreaction due to growth
of metric perturbations\cite{SMP}. However, we shall see that this
is unnecessary here since long-wavelength modes do not grow
resonantly. Further, the Hartree approximation misses rescattering
effects \cite{pre2}, which may be important at the final stage of 
preheating. 
We leave both of these issues for future work. 

\begin{figure}
\begin{center}
\singlefig{9cm}{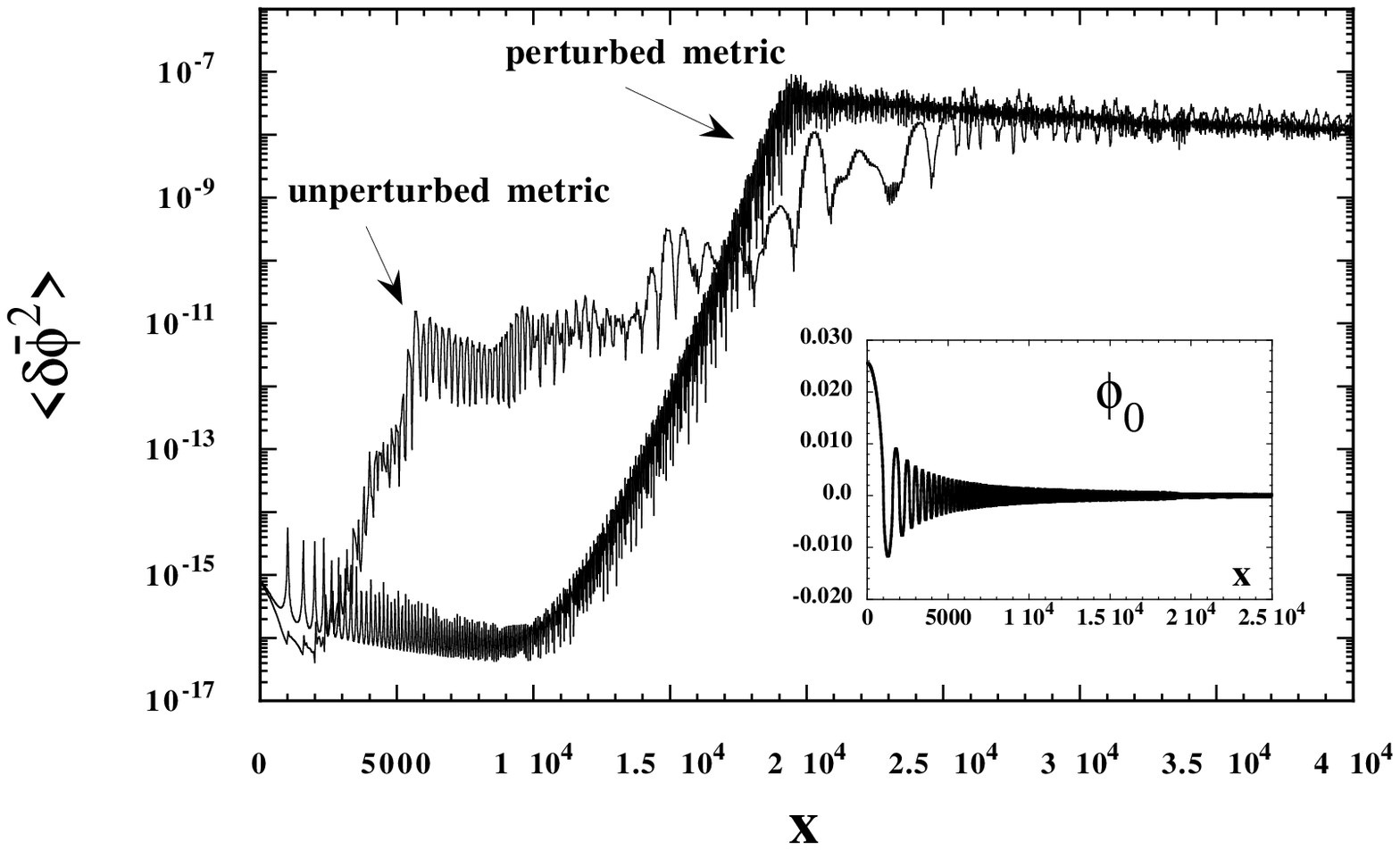}
\begin{figcaption}{Fig1}{9cm}
Time evolution of the inflaton variance  $\langle \delta
\bar{\phi}^2 \rangle \equiv \langle \delta \phi^2 \rangle/ m_{\rm
pl}^2$ for $\xi=-70$ and $\lambda=10^{-12}$,  both for the
perturbed and unperturbed ($\Phi_k = 0$) metrics. In the
unperturbed metric, the variance initially grows due to the
enhancement of small-$k$ modes. However, when metric perturbations
are included, growth only occurs after $x \approx 10^4$  due to
amplification of sub-Hubble modes. {\bf Inset:} Evolution of the
condensate, $\phi_0$, vs $x$, for $\xi=-70$. 
\end{figcaption}
\end{center}
\end{figure}

In the minimally coupled multi-field case, large resonance
parameters may lead to suppression of the long-wavelength modes of
the preheating fields  during inflation \cite{mpre3}. Through the
constraint equation [c.f. Eq.~(\ref{B5})], this removes the
resonance in the long-wavelength modes of the metric
perturbations, $\Phi_k$.

For large $|\xi|$ the same effect might be expected. In fact, for
large {\em negative} $\xi$, the opposite occurs, and it is the
short-wavelength modes which are suppressed relative to the $k
\sim 0$ modes, leading to a red spectrum at the start of
preheating \cite{SH}. \footnote{The modes in de Sitter space evolve
as Hankel functions with order $\nu = (9/4 - m^2/H^2 -
12\xi)^{1/2}$, which is positive  when $\xi \ll -1$. For
suppression to exist, complex $\nu$ is required.}

Intuitively one sees that since the scalar curvature during
inflation is  $R \approx -\lambda \phi_0^2/\xi$ for $|\xi| \gg 1$
by Eqs.~$(\ref{B9})$ and $(\ref{B8})$, 
the frequency $\omega_k^2 \equiv k^2/a^2+
3\lambda\phi_0^2+\xi R$ on the l.h.s. of Eq.~$(\ref{B6})$ is
$\omega_k^2 \approx k^2/a^2+2\lambda\phi_0^2$. This indicates that
the suppression mechanism is {\it absent} during inflation and
cannot be responsible for removing the resonances of preheating.

Before analyzing the evolution of the fluctuations during
preheating numerically, we analytically estimates these
quantities. Introducing new variables
$\hat{\Phi}_k=\Phi_k+\delta{F}_k/2F,~\hat{\Psi}_k=\Psi_k-\delta{F}_k/2F$,
Eqs.~$(\ref{B4})$-$(\ref{B60})$ yield
\begin{eqnarray}
\hat{\Phi}_k=\hat{\Psi}_k,~~~\delta\phi_k=\frac{2F^{3/2}}{\kappa^2
a\dot{\phi}_0 E} (a\sqrt{F}\hat{\Psi}_k)^{\cdot}, 
\label{B12}
\end{eqnarray}
\begin{eqnarray}
& &\ddot{\hat{\Psi}}_k+\left(4H+\frac{3\dot{F}}{2F}\right)
\dot{\hat{\Psi}}_k \nonumber \\
&+& \left(\frac{k^2}{a^2}-\frac{\kappa^2\dot{\phi}_0^2}{2F}
-\frac{3\dot{F}^2}{4F^2}
+\frac{\kappa^2\lambda\phi_0^4}{4F}\right) \hat{\Psi}_k \nonumber \\
&=&
\frac34 \left(\frac{2\ddot{F}}{F}+3H\frac{\dot{F}}{F}
-\frac{3\dot{F}^2}{2F^2}+\frac{2\kappa^2\dot{\phi}_0^2}{3F}
\right)\frac{\delta F_k}{F} \nonumber \\
&-&
\kappa^2\left[\xi R \phi_0+\lambda\phi_0^3+\frac32 \dot{\phi}_0
\left(H+\frac{\dot{F}}{2F}\right)\right] \frac{\delta\phi_k}{F},
\label{B36}
\end{eqnarray}
where $E=1-(1-6\xi)\xi\kappa^2\phi_0^2$.
Note that $\hat{\Phi}_k$ and $\hat{\Psi}_k$ are conformally transformed
potentials which are derived in the Einstein frame\cite{MK}.
The relation $(\ref{B12})$ clearly shows
the link between inflaton fluctuations and  metric perturbations.
Eliminating $\delta F_k/F$ and $\delta \phi_k/F$ terms in 
Eq.~$(\ref{B36})$ by using the relation $(\ref{B12})$, 
we obtain
\begin{eqnarray}
\frac{d^2u_k}{d\eta^2}+\left(k^2-\frac{1}{z}
\frac{d^2 z}{d\eta^2}\right)u_k=0,
\label{B14}
\end{eqnarray}
where $\eta \equiv \int a^{-1}dt$ is  conformal time, and
\begin{eqnarray}
u_k \equiv \frac{ F^{3/2}}{\dot{\phi}_0 \sqrt{E}}
\hat{\Psi}_k,~~~ z \equiv
\frac{(a\sqrt{F})^{\cdot}}{a^2\dot{\phi}_0 \sqrt{E}} \,.
\label{B15}
\end{eqnarray}
In the long-wave limit, $k \rightarrow 0$, Eq.~$(\ref{B14})$ is
easily integrated to give
\begin{eqnarray}
u_k =z \left(c_1+c_2 \int \frac{d\eta}{z^2} \right),
\label{B16}
\end{eqnarray}
where $c_1$ and $c_2$ are constants. Making use of
Eqs.~$(\ref{B12})$, $(\ref{B15})$, and $(\ref{B16})$, the
long-wavelength solutions are
\begin{eqnarray}
\Phi &=& -\left(\frac{1}{aF}\right)^{\cdot}
\left(c_1-2c_2\int aFdt \right)+2c_2, \\
\label{B17}
\Psi &=& \frac{\dot{a}}{a^2F}
\left(c_1-2c_2\int aFdt \right)+2c_2, \\
\label{B18}
\delta\phi &=& -\frac{\dot{\phi}_0}{aF}
\left(c_1-2c_2\int aFdt \right).
\label{B32}
\end{eqnarray}
In the present model, the inflationary period ends when the value
of $|\xi|\kappa^2\phi_0^2$ decreases to of order
unity\cite{FU,nonminimalpre}. This means that
$F=1-\xi\kappa^2\phi_0^2$ is of order unity during preheating, and
does not change significantly. Since numerical calculations
indicate that the scale factor roughly evolves as a power-law in
the oscillating stage of inflaton, we can expect from Eqs.~(2.18)
and $(\ref{B17})$ that super-Hubble metric perturbations do not
grow significantly during preheating, which restricts the
enhancement of the inflaton fluctuations for small-$k$ modes by
Eq.~$(\ref{B12})$. In fact, Eq.~$(\ref{B32})$ implies that 
the difference between the minimally coupled case only appears 
in the $F$ term, which approaches unity with the decrease of 
$\phi_0$.

This contradicts the earlier work\cite{nonminimalpre} neglecting
metric perturbations, which pointed out that the $\xi R$ term in
the l.h.s. of Eq.~$(\ref{B6})$ leads to the growth of low momentum
modes for the strong coupling case. In the perturbed metric case,
this term is counteracted by the last term
in Eq.~$(\ref{B6})$. Actually, eliminating the 
$\ddot{\Psi}_k$ term in the r.h.s. of Eq.~$(\ref{B6})$ by using
Eq.~$(\ref{B60})$ gives rise to the $\xi R\delta\phi_k$ term
in the r.h.s. of $(\ref{B6})$, which suppresses the resonance 
due to the curvature term.

In the next section, we numerically confirm this lack of
resonance.

 \section{Removal of the  super-Hubble resonance}

We numerically solved the perturbation  equations
$(\ref{B4})$-$(\ref{B6})$  which are coupled to the  background
equations $(\ref{B7})$-$(\ref{B9})$ through the fluctuation
integrals (2.11) which mediate backreaction effects. We started
integrating at the onset of preheating, with  initial values of
the inflaton oscillations given by
\begin{eqnarray}
\phi_0 (0)= \left[\frac{\sqrt{(1-24\xi)(1-8\xi)}-1}
{16\pi(1-6\xi)|\xi|} \right]^{1/2} m_{\rm pl},
\label{B19}
\end{eqnarray}
which is determined by the end of slow-roll inflation
\cite{nonminimalpre}. We choose the conformal vacuum state for the
initial field fluctuations:  $\delta\phi_k=1/\sqrt{2\omega_k(0)}$
and $\delta\dot{\phi}_k=[-i\omega_k(0)-H(0)]\delta\phi_k$. The
gauge-invariant potentials $\Phi_k$ and $\Psi_k$  are then
determined when the evolution of the scalar field is known (see
Eq.~(34) in ref.~\cite{Hwang}).

In the minimally coupled case, $\xi=0$, and neglecting metric
perturbations,  there is a  resonance band in the narrow,
sub-Hubble, range\cite{structure}:
\begin{eqnarray}
3/2<\bar{k}^2<\sqrt{3},
\label{B20}
\end{eqnarray}
with $\bar{k}^2 \equiv k^2/(\lambda \tilde{\phi}_0^2(0))$, where
$\tilde{\phi}_0(0)$ is the initial amplitude of inflaton. In that
case, the growth of field fluctuations finally stops due to the
backreaction of created particles, and the final variance is
$\langle\delta\phi^2\rangle \approx 10^{-7}m_{\rm pl}^2$
\cite{KT1,nonminimalpre}.

In the perturbed spacetime with $\xi=0$, although metric
perturbations assist the enhancement of field fluctuations, only
sub-Hubble modes of field and metric perturbations are amplified
unless mode-mode coupling is taken into account as in
ref.~\cite{mpre5}. We find numerically  that the final variance is
almost the same both in the perturbed and unperturbed metric. We
warn that this is not a generic feature of preheating since in the
multi-field case, alternative channels exist for resonant decay
\cite{mpre2,mpre4}.

\begin{figure}
\begin{center}
\singlefig{9cm}{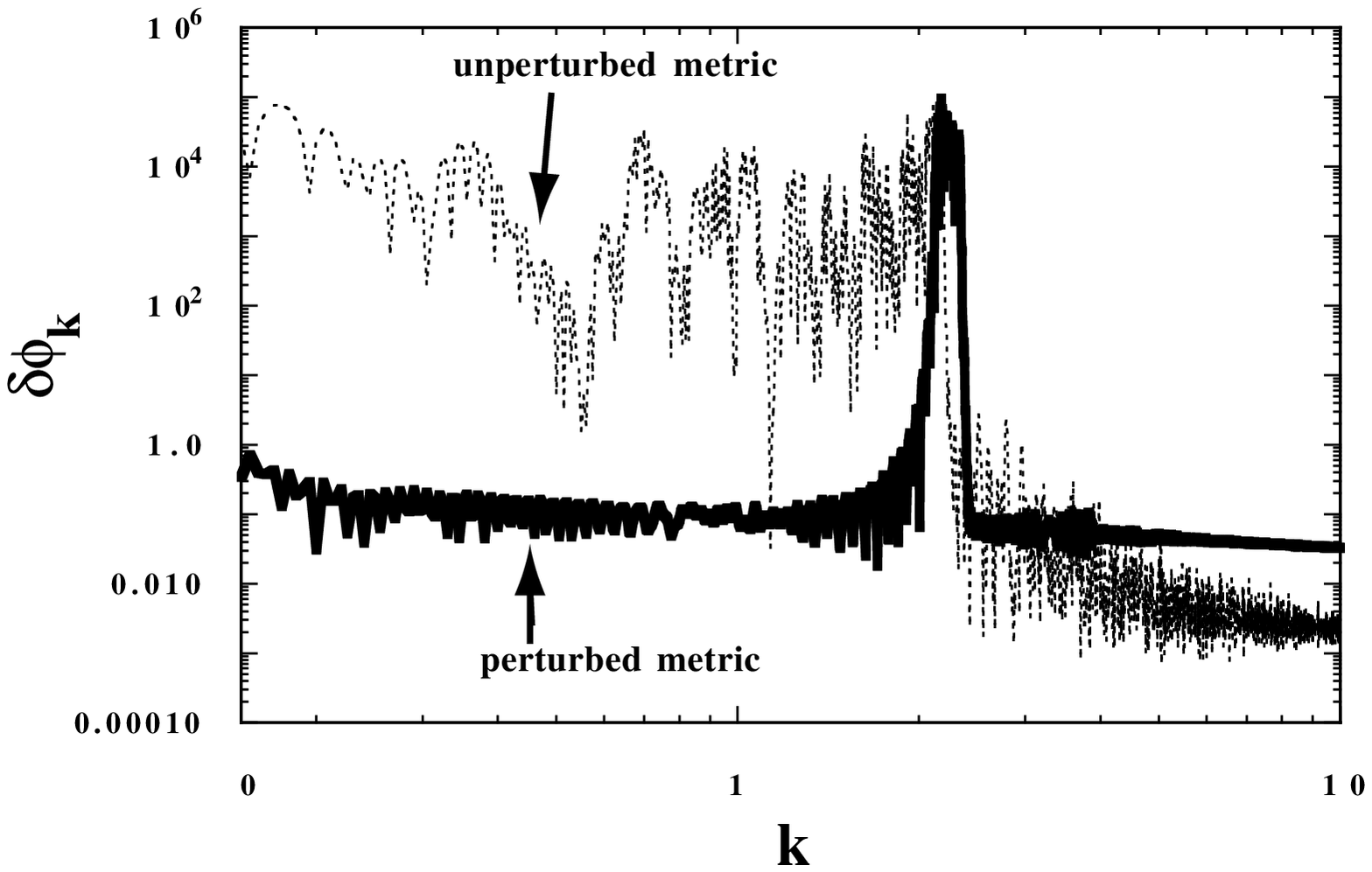}
\begin{figcaption}{Fig2}{9cm}
The spectrum of field fluctuations: $\delta\phi_k$ vs $k$, for
$\xi=-100$ for both the perturbed and unperturbed metrics. When
metric perturbations are included, the super-Hubble resonance
disappears and is replaced by a single, sub-Hubble band.
\end{figcaption}
\end{center}
\end{figure}

In the non-minimally coupled case with  metric perturbations
neglected (the {\em rigid} case), low momentum $(k \sim 0$) field
modes undergo resonant amplification during preheating for
$\xi~\lsim~-10$\cite{nonminimalpre,nonminimalpre2}. With
increasing $|\xi|$, the duration during which these low momentum
modes are amplified becomes gradually longer, and they dominate
the  final variance.

When the spacetime metric is perturbed, as it must be for
consistency, these metric fluctuations on the r.h.s of
Eq.~$(\ref{B6})$ lead to a very different picture. In Fig.~(1) we
show the evolution of the variance $\langle\delta\phi^2\rangle$
for $\xi=-70$, both in the unperturbed and perturbed spacetimes.
With metric perturbations ignored, small-$k$ field fluctuations
exhibit strong growth and the variance grows for $x~\lsim~6000$,
where $x \equiv \sqrt{\lambda}\eta  m_{\rm pl}$ is the natural
dimensionless conformal time of the system. For $x~\gsim~6000$ the
effect of the non-minimal coupling becomes insignificant due to
the decrease in the scalar curvature. Sub-Hubble modes then
dominate the variance, as in the minimally coupled case, which
continues to grow until $x \sim 2.5 \times 10^4$.

When metric perturbations are included, the variance
$\langle\delta\phi^2\rangle$ does not grow  initially, but
oscillates with much higher frequency. This implies that the
small-$k$ modes are not amplified, as expected from our earlier,
analytical, discussion. 

The absence of the super-Hubble resonance can be seen in Fig. (2) 
where we show the spectrum of $\delta\phi_k$ at constant time. In the
perturbed metric, there is only a  small resonance band around
$k/(\sqrt{\lambda}\phi_0(0)) \simeq 2.2$, while in the unperturbed
metric, the negative coupling instability creates a resonant band
that contains all modes with $k/(\sqrt{\lambda}\phi_0(0))~\lsim~2$.

The effect of this change to the resonance structure is clearly
visible in Fig.~(3) where we plot a cosmological mode
$(k/(\sqrt{\lambda}\phi_0(0)) = 10^{-25})$ of the metric
perturbation $\Psi_k$, which shows no growth, and a sub-Hubble
mode, which grows resonantly until backreaction ends its
amplification.

\begin{figure}
\begin{center}
\singlefig{9cm}{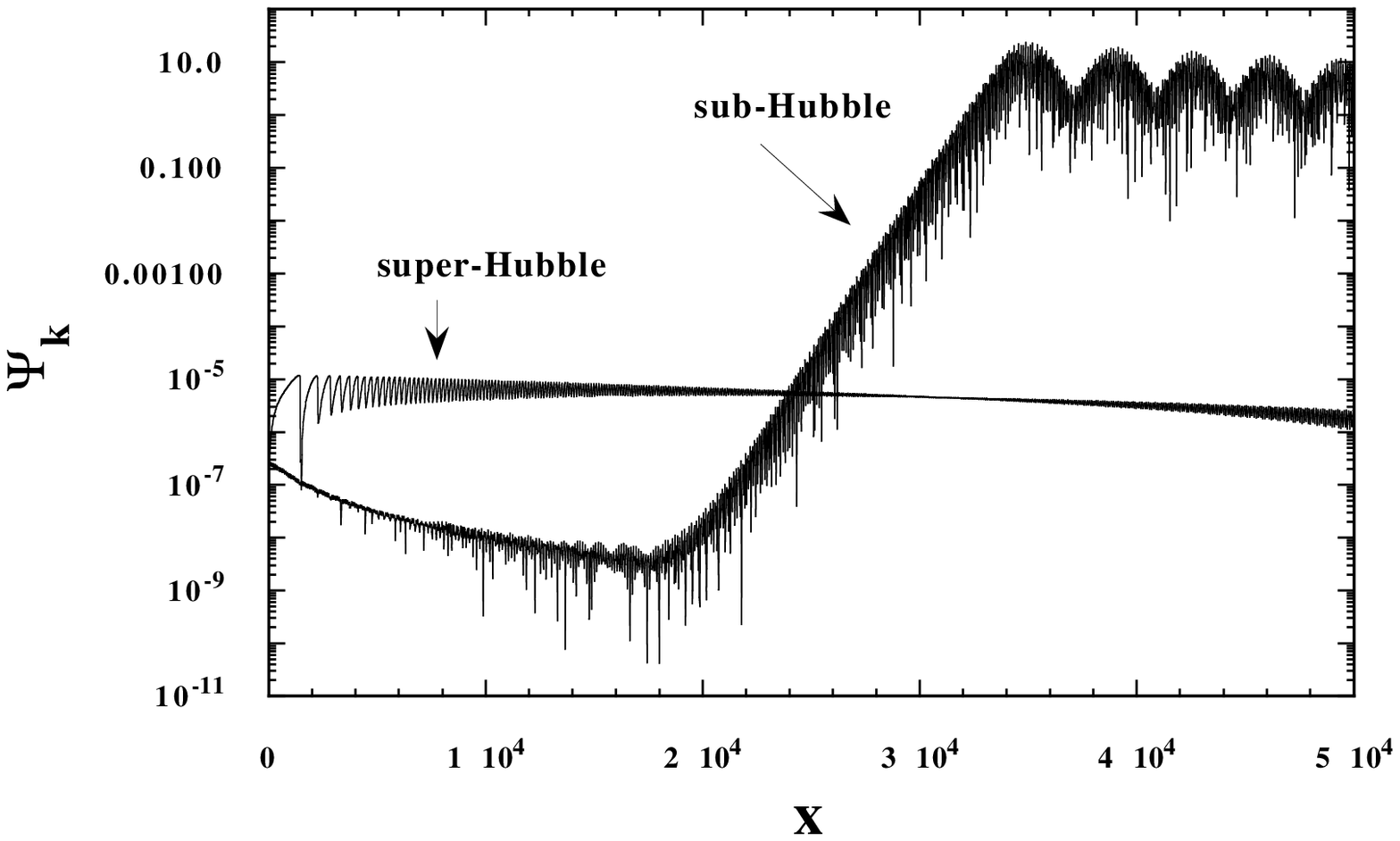}
\begin{figcaption}{Fig3}{9cm}
The
evolution of the metric fluctuation $\Psi_k$ for the super-Hubble
mode $k/(\sqrt{\lambda}\phi_0(0)) =10^{-25}$ and the sub-Hubble
mode $k/(\sqrt{\lambda}\phi_0(0))=2.2$ with $\xi=-100$. The
evolution of $\Phi_k$ shows a similar behavior. We find that
super-Hubble metric and field fluctuations  do not grow during
preheating in this model.
\end{figcaption}
\end{center}
\end{figure}

We found numerically that the final variances of field
fluctuations when $\Phi_k$ is included are typically smaller, by one or two
magnitudes, than in the rigid spacetime  for
$|\xi|~\gsim~100$, although they are almost the same for
$|\xi|~\lsim~100$. This is understandable, since field
fluctuations without $\Phi_k$ for $|\xi|~\gsim~100$ are dominated
by low momentum modes and $\langle\delta\phi^2\rangle$ grows to of
order $\phi_0^2$\cite{nonminimalpre}, while in the perturbed
spacetime, resonance stops before $\langle\delta\phi^2\rangle$
reaches $\phi_0^2$, as in the case of $\xi=0$.

The lack of strong metric preheating on cosmological scales
persists in the strong coupling regime $|\xi|~\gsim~10^3$
considered by Fakir and Unruh. The predictions of large scale
metric perturbations produced in the inflationary
epoch\cite{MS,KF} are therefore not modified in the present model,
as long as another scalar field coupled non-gravitationally to the
inflaton is not introduced.

 \section{Discussion and conclusions}

We have studied preheating in a non-minimal, chaotic, inflation
scenario. Our main result is that consistent inclusion of metric
perturbations is crucial and has a powerful and unexpected  
impact on the evolution of long-wavelength fluctuations 
during preheating, removing the exponential growth of these 
modes that one finds in the absence of metric perturbations.

This is in strong contrast to the multi-field case where the
negative specific heat of gravity tends to enhance any resonances
that exist in their absence \cite{mpre2,mpre4}. In addition, large
negative $\xi$ causes anti-suppression of long-wavelength modes,
relative to short wavelength modes during inflation, providing an
alternative way of avoiding the suppression mechanisms proposed in
\cite{mpre3}.

The removal of the resonances is intimately related to the
existence of  gauge-invariant conserved quantities in the
long-wavelength limit:
\begin{eqnarray}
\zeta \equiv \Psi-\frac{\dot{a}^2}{a^2F [\dot{a}/(aF)]^{\cdot}}
\left( \Psi+\frac{a}{\dot{a}} \dot{\Psi} \right)=2c_2,
\label{B21}
\end{eqnarray}
which is obtained from Eq.~$(\ref{B17})$.

A simple way to appreciate the lack of super-Hubble resonances is
to recognize that the present system is conformally equivalent to
a minimally coupled, single scalar field, model with potential
$\hat{V}(\phi)\equiv V(\phi)/(1-\xi\kappa^2\phi^2)^2$ and
$|\xi|\kappa^2\phi^2~\sim~1$ at the end of inflation. Hence it is
not surprising that the resonance structure is similar to that of
the minimally coupled case \cite{structure}. However, this 
equivalence is missing if metric perturbations are not included.

On the other hand, on sub-Hubble scales, we find that both field
and metric fluctuations are excited during preheating as in the
minimally coupled case. This may result in an important
cosmological consequences such as the production of primordial
black holes \cite{mpre2}. There may also be interesting
implications for non-thermal symmetry restoration due to the
change in the time evolution of the variance.

Although we have restricted ourselves to a  non-minimally coupled
scalar field in Einstein gravity, the evolution of metric
perturbations can be analyzed in a unified manner in  generalized
Einstein theories, which include the higher-curvature,
Brans-Dicke, and induced gravity theories\cite{Hwang}. As long as
the system we consider is a single-field model, and stress-energy
is conserved, super-Hubble metric perturbations will not be
enhanced during preheating, because conserved quantities,  which
generalize $\zeta$ in Eq.~$(\ref{B21})$, exist.

However, introducing a coupling such as the standard
$g^2\phi^2\chi^2/2$, will violate the conservation of the quantity
$\zeta$ for certain values of $g$ and $\xi$, as occurs in the minimally
coupled multi-field case \cite{mpre4}.

Finally it is of great interest to examine the realistic
multi-field case in known classes of inflationary models, because
we can constrain the inflaton potential in terms of the distortion
of the CMB spectrum due to parametric resonance in preheating.
This requires a complete study of backreaction including metric
perturbations, which we leave to future work.

\section*{ACKOWLEDGEMENTS}
We  thank Roy Maartens, Kei-ichi Maeda, Takashi Torii and David
Wands for useful discussions. 

\end{document}